\documentstyle[12pt]{article}
\title{\bf Towards the Realistic Gluodynamics String. Perturbative
Gluons' Contribution to the String Effective Action}
\author{D.V.ANTONOV \thanks{E-mail addresses:
antonov@pha2.physik.hu-berlin.de, antonov@vxitep.itep.ru, supported
by Graduiertenkolleg {\it Elementarteilchenphysik}, Russian
Fundamental Research Foundation, Grant No.96-02-19184, DFG-RFFI,
Grant 436 RUS 113/309/0 and by the Intas, Grant No.94-2851.}
\\
{\it Institute of Theoretical and Experimental Physics}\\
{\it B.Cheremushkinskaya 25, 117218, Moscow, Russia}\\
{\it and}\\
{\it Institut f\"ur Elementarteilchenphysik,
Humboldt-Universit\"at,}\\
{\it Invalidenstrasse 110, D-10115, Berlin, Germany}\\}
\date{}
\begin{document}
\maketitle
\vspace{1mm}
\centerline{\bf {Abstract}}
\vspace{3mm}
Perturbation theory in the nonperturbative QCD vacuum and the
non-Abelian Stokes theorem, representing a Wilson loop in the $SU(2)$
gluodynamics as an integral over all the orientations in colour
space, are applied to derivation of the correction to the string
effective action, obtained in Ref. 1. This correction is due to
accounting in the lowest order of perturbation theory for the
interaction of perturbative gluons with the string world sheet. It
occurs that this interaction affects only the coupling constant of
the rigidity term, while its contribution to the string tension of
the Nambu-Goto term vanishes. The obtained correction to the rigid
string coupling constant multiplicatively depends on the spin of the
representation of the Wilson loop under consideration, the QCD
coupling constant and a certain path integral, which includes the
background Wilson average.

\vspace{6mm}
{\large \bf 1. Introduction}

\vspace{3mm}
Recently, a new approach to the gluodynamics string was
suggested$^{1}$. Within this approach one considers the Wilson
average written through the non-Abelian Stokes theorem$^{2,3}$ and
the cumulant expansion$^{3,4}$ as a statistical weight in the
partition function of some effective string theory. The action of
this string theory may be then obtained via expansion of the Wilson
average in powers of two small parameters,
$\left(\alpha_sF_{\mu\nu}^a(0)F_{\mu\nu}^a(0)\right)^
{\frac{1}{2}}T_g^2$ and $\left(\frac{T_g}{r}\right)^2$, where
$T_g\simeq 0,2 fm$ is the correlation length of the vacuum$^{5,6}$,
and $r\simeq 1 fm$ is the size of the Wilson loop in the confining
regime$^{7}$. This yields the so-called curvature expansion for the
gluodynamics string effective action. In Ref. 1 only the first
nonvanishing terms of this expansion, corresponding to the lowest,
second, order in the former parameter were accounted for, which
corresponds to the so-called bilocal approximation$^{3, 8-11}$, and
the expansion up to the terms of the third order in the latter
parameter was elaborated out. The first two terms of this expansion
read as follows

$$S_{biloc.}=\sigma\int d^2\xi\sqrt{g}+\frac{1}{\alpha_0}\int
d^2\xi\sqrt{g}g^{ij}\left(\partial_it_{\mu\nu}\right)\left(\partial_j
t_{\mu\nu}\right), \eqno (1)$$
where

$$\sigma=4T_g^2\int d^2zD\left(z^2\right) \eqno (2)$$
is the string tension of the Nambu-Goto term, and

$$\frac{1}{\alpha_0}=\frac{1}{4}T_g^4\int d^2zz^2\left(2D_1\left(z^2
\right)-D\left(z^2\right)\right) \eqno (3)$$
is an inverse bare coupling constant of the rigidity term, while the
terms of the third order in $\left(\frac{T_g}{r}\right)^2$ contain
higher derivatives of the induced metric $g_{ij}$ and/or of the
extrinsic curvature tensor $t_{\mu\nu}$ w.r.t. world sheet
coordinates, and we shall not quote them here (see Ref. 1 for the
details)\footnote{From now on we shall use for the world sheet
indices the letters from the middle of the Latin alphabet, $i, j,
k,...,$ in order not to confuse them with the colour indices $a, b,
c,...$. We shall also keep the standard notation "$g$" for the QCD
coupling constant and would like to prevent the reader from confusing
it with the determinant of the induced metric tensor,
denoted by the same letter.}. In Eqs. (2) and (3) $D$ and $D_1$ stand
for two renormalization group invariant coefficient functions,
parametrizing the gauge-invariant bilocal correlator of gluonic field
strength tensors$^{3, 8-10}$, and it is worth noting that since the
nonperturbative parts of these functions are related to each other as
$\mid D_1\mid\simeq\frac{1}{3}D$ according to lattice data$^{6}$, the
inverse rigid string coupling constant (3) is negative, which
according to Ref. 12 agrees with the mechanism of confinement, based
on the dual Meissner effect$^{13}$ (see discussion in Ref. 1). This
result confirms that the Method of Vacuum Correlators, developed in
Refs. 3 and 8-11 (see also Refs. therein), provides us with the
consistent description of the confining gluodynamics vacuum. The
approach suggested in Ref. 1 enables one to express all the coupling
constants of the terms emerging in the string action in higher orders
of the curvature expansion through the gauge-invariant correlators of
gluonic field strength tensor only.

Notice also, that in Ref. 14 action (1) was applied to the derivation
of the correction to the Hamiltonian of the QCD string with quarks,
which was obtained in Ref. 15, arising due to the rigidity term, with
the help of which a rigid string-induced term in the Hamiltonian of
the relativistic quark model was then evaluated for the case of large
masses of a quark and antiquark.

However, it should be emphasized that the curvature expansion
describes only the pure nonperturbative content of the gluodynamics
string theory. As it was explained in Ref. 10, in order to get the
correct spectrum of the open bosonic string and the exponential
growth of the multiplicity of string states, which is necessary for
ensuring the property of duality of amplitudes, contained in the
Veneziano formula, one must account for the perturbative gluons
interacting with the string, which can be done in the framework of
the perturbation theory in the nonperturbative QCD vacuum$^{9}$
(see discussion in Ref. 1).

In this letter, we shall take this interaction into account in the
lowest order of perturbation theory and obtain the corresponding
correction to the action (1). To this end, one needs to integrate
over perturbative fluctuations in the expression for the Wilson
average written through the non-Abelian Stokes theorem. This
procedure is, however, looks rather difficult to be elaborated out in
the case when one makes use of the version of the non-Abelian Stokes
theorem, suggested in Refs. 2 and 3, due to the path-ordering, which
is remained in the expression for the Wilson loop after rewriting it
as a surface integral. In what follows, in order to get rid of it, we
shall exploit another version of the non-Abelian Stokes theorem,
which was proposed in Ref. 16, where the path-ordering was replaced
by the integration over an auxiliary field from the
$SU\left(N_c\right)/\left[U(1)\right]^{N_c-1}$ coset space. For
simplicity we shall consider the $SU(2)$-case, when this field is a
unit three-vector $\vec n$, which characterizes the instant
orientation in the colour space, and the non-Abelian Stokes theorem
takes a remarkably simple form. Integration over perturbative
fluctuations then yields the interaction of the elements of the
string world sheet via the nonperturbative gluonic exchanges. In
other words, we arrive at a theory of the open nonperturbative
strings between dynamical world sheet elements, which provides us
with the quantitative description of the intuitive picture, described
in the previous paragraph. Finally, due to the short rangeness of
this nonperturbative interaction between world sheet elements, it
occurs possible to perform the $\vec n$-averaging and extract all the
remnant dependence on the background fields in the form of the Wilson
average standing under a certain path integral, so that the
dependence on the world sheet elements decouples and may be evaluated
explicitly, which yields a correction to the rigidity term, while the
string tension of the Nambu-Goto term does not acquire any
corrections and keeps its pure nonperturbative value (2). All the
points, mentioned above, will be worked out in the next Section.

The main results of the letter are summarized in the Conclusion.

\vspace{6mm}
{\large \bf 2. An Action of the Gluodynamics String Including
Perturbative Gluons' Contributions}

\vspace{3mm}
The statistical weight of the effective string theory, we are going
to derive, is the Wilson average in the $SU(2)$ gluodynamics $\left<
W(C)\right>=\left<tr~P~\exp\left(ig\oint\limits_C^{} dx_\mu
A_\mu^at^a\right)\right>$, which after rewriting it as a surface
integral by virtue of the non-Abelian Stokes theorem, suggested in
Ref. 16, splitting the total field $A_\mu^a$ into the background
$B_\mu^a$ and the perturbative fluctuations $a_\mu^a,
A_\mu^a=B_\mu^a+ a_\mu^a$, and making use of the background field
formalism$^{9,17}$, takes the form

$$\left<W(C)\right>=\int DB_\mu^a Da_\mu^a D\vec n \exp\Biggl[\int dx
\Biggl[-\frac{1}{4}\left(F_{\mu\nu}^a\right)^2+\frac{1}{2}a_\mu^a
D_\lambda^{ab}D_\lambda^{bc}a_\mu^c+a_\nu^aD_\mu^{ab}F_{\mu\nu}^b+$$

$$+\frac{iJ}{2}\int
d\sigma_{\mu\nu}n^a\left(-g\left(F_{\mu\nu}^a+2D_\mu^{ab}a_\nu^b\right)+
\varepsilon^{abc}\left(D_\mu\vec n{\,}\right)^b\left(D_\nu\vec
n{\,}\right)^c\right)\Biggr]\Biggr], \eqno (4)$$
where $F_{\mu\nu}^a=\partial_\mu B_\nu^a-\partial_\nu
B_\mu^a+g\varepsilon^{abc}B_\mu^bB_\nu^c$ is a strength tensor of the
background field,
$D_\mu^{ab}=\delta^{ab}\partial_\mu-g\varepsilon^{abc}B_\mu^c$ is the
corresponding covariant derivative, and
$J=\frac{1}{2},1,\frac{3}{2},...$ is the spin of the representation
of the Wilson loop under consideration. In what follows we shall be
interested only in the effects of the lowest order of perturbation
theory, so that on the R.H.S. of Eq. (4) we have replaced the full
covariant derivative which should stand in the last term (the
so-called Wess-Zumino term) by the background one and omitted the
ghost term, the term, describing the interaction of two perturbative
gluons with the background field strength tensor and with the string
world sheet, and the terms which describe the interaction of three
and four perturbative gluons. Integration over the perturbative
fluctuations in Eq. (4) is Gaussian and yields

$$\left<W(C)\right>=\int DB_\mu^a D\vec n \exp\Biggl[-\frac{1}{4}\int
dx\left(F_{\mu\nu}^a\right)^2+\frac{iJ}{2}\int
d\sigma_{\mu\nu}\varepsilon^{abc}n^a\left(D_\mu\vec n{\,}\right)^b
\left(D_\nu\vec
n{\,}\right)^c\Biggr]\cdot$$

$$\cdot\exp\Biggl(-\frac{iJg}{2}\int
d\sigma_{\mu\nu}n^aF_{\mu\nu}^a\Biggr)\exp\Biggl[-\frac{1}{2}\int dxdy
\Biggl[D_\mu^{ba}\left(iJgT_{\mu\nu}^a(x)+F_{\mu\nu}^a(x)\right)\Biggr]
\int\limits_0^\infty ds\int (Dz)_{xy}e^{-\int\limits_0^s\frac{\dot
z^2}{4}d\lambda}\cdot$$

$$\cdot\Biggl[P~\exp\Biggl(ig\int\limits_0^s
d\lambda\dot z_\alpha
B_\alpha\Biggr)\Biggr]^{bc}\Biggl[D_\rho^{cd}\left(
iJgT_{\rho\nu}^d(y)+F_{\rho\nu}^d (y)\right)\Biggr]\Biggr], \eqno
(5)$$
where $T_{\mu\nu}^a(x)\equiv\int
d^2\xi\varepsilon^{ij}\left(\partial_ix_\mu(\xi)\right)
\left(\partial_jx_\nu(\xi)\right)n^a\left(x(\xi)\right)
\delta\left(x-x(\xi)\right)$ is the colour vorticity tensor current.
Making use of the formula$^{4}$

$$\left<e^PQ\right>=\left(\exp\left(\sum\limits_{n=1}^\infty
\frac{1}{n!}\left<\left<
P^n\right>\right>\right)\right)\left(\left<Q\right>+
\sum\limits_{k=1}^\infty\frac{1}{k!}
\left<\left< P^kQ\right>\right>\right),$$
where $P$ and $Q$ stand for two statistically dependent commuting
quantities, and $\left<\left<...\right>\right>$ denotes the so-called
cumulants, i.e. irreducible correlators, we get in the lowest order
of the cumulant expansion the following correction to the string
effective action (1)

$$\Delta S=-\frac{J^2g^2}{2}\int
dxdy\Biggl<\left(D_\mu^{ba}T_{\mu\nu}^a(x)\right)\int\limits_0^\infty
ds\int\left(Dz\right)_{xy}e^{-\int\limits_0^s\frac{\dot
z^2}{4}d\lambda}\cdot$$

$$\cdot\Biggl[P~\exp\Biggl(ig\int\limits_0^s
d\lambda\dot z_\alpha
B_\alpha\Biggr)\Biggr]^{bc}D_\rho^{cd}T_{\rho\nu}^d(y)\Biggr>_{\vec
n{\,}, B_\mu^a}, \eqno (6)$$
where

$$\left<...\right>_{\vec n{\,}}\equiv\int D\vec
n{\,}\left(...\right)\exp\left(\frac{iJ}{2}\int
d\sigma_{\mu\nu}\varepsilon^{abc}n^a\left(D_\mu\vec
n{\,}\right)^b\left(D_\nu \vec n{\,}\right)^c\right),$$

$$\left<...\right>_{B_\mu^a}\equiv\int
DB_\mu^a\left(...\right)\exp\left(-\frac{1}{4}\int
dx\left(F_{\mu\nu}^a\right)^2\right),$$
and during the derivation of
Eq. (6) we have omitted the term

$$\exp\left[-\frac{1}{2}\int
dxdy\left(D_\mu^{ba}F_{\mu\nu}^a(x)\right)\int\limits_0^\infty ds\int
\left(Dz\right)_{xy}e^{-\int\limits_0^s\frac{\dot
z^2}{4}d\lambda}\left[P~\exp\left(ig\int\limits_0^s d\lambda\dot
z_\alpha
B_\alpha\right)\right]^{bc}D_\rho^{cd}F_{\rho\nu}^d(y)\right]$$
on the R.H.S. of Eq. (5), which does not yield any contributions to
the string action due to the lack of coupling with the world sheet
and therefore may be absorbed into the measure $DB_\mu^a$.
Integrating in Eq. (6) by parts, we arrive in the lowest order of
perturbation theory at the following formula

$$\Delta S=-\frac{J^2g^2}{2}\int d^2\xi\int
d^2\xi^\prime\varepsilon^{ij}\varepsilon^{kl}\left(\partial_ix_\mu\right)
\left(\partial_jx_\nu\right)(\partial_kx_\rho^\prime)\left(
\partial_lx_\nu^\prime\right)\cdot$$

$$\cdot\left<\left<n^b\left(x\right)n^c\left(
x^\prime \right)\right>_{\vec n{\,}}\frac{\partial^2}{\partial x_\mu
\partial x_\rho^\prime}\int\limits_0^\infty
ds\int\left(Dz\right)_{xx^\prime}e^{-\int\limits_0^s\frac{\dot
z^2}{4}d\lambda}\left[P~\exp\left(ig\int\limits_0^s d\lambda\dot
z_\alpha B_\alpha\right)\right]^{bc}\right>_{B_\mu^a}, \eqno (7)$$
where $x_\mu\equiv x_\mu\left(\xi\right)$, and $x_\mu^\prime\equiv
x_\mu\left(\xi^\prime\right)$.

Since $e^{-\int\limits_0^s\frac{\dot
z^2}{4}d\lambda}P~\exp\left(ig\int\limits_0^s d\lambda\dot z_\alpha
B_\alpha\right)$ is the statistical weight of a perturbative gluon,
propagating from the point $x^\prime$ to the point $x$ along the
trajectory $z_\alpha$ during the proper time $s$, it is the region
where $s$ is small, which mainly contributes to the path integral on
the R.H.S. of Eq. (7). This means that the dominant contribution to
$\Delta S$ comes from those $x_\mu$'s and $x_\mu^\prime$'s, which are
very close to each other, which is in the line with the curvature
expansion, where $\mid x^\prime-x\mid\le T_g\ll r$. Within this
approximation, one gets

$$\left<n^b\left(x\right)n^c\left(x^\prime\right)\right>_{\vec n{\,}}
\simeq\frac{\delta^{bc}}{3}\int D\vec n{\,}\exp\left(\frac{iJ}{2}
\int d\sigma_{\mu\nu}\varepsilon^{def}n^d\left(D_\mu\vec
n{\,}\right)^e\left(D_\nu\vec n{\,}\right)^f\right). \eqno (8)$$
It is worth noting, that the integral on the R.H.S. of Eq. (8) is a
functional of the world sheet as a whole (it is independent of
$x_\mu(\xi)$), and therefore may be also referred to the measure
$DB_\mu^a$.

Hence, as it was announced in the Introduction, we see that
expression (7) for the correction to the string effective action
(1) due to the perturbative gluons takes the form of the interaction
of two elements of the world sheet, $d\sigma_{\mu\nu}(\xi)$ and
$d\sigma_{\rho\nu}\left(\xi^\prime\right)$, via the nonperturbative
gluonic string.

In order to extract explicitly the dependence on the points $x$ and
$x^\prime$ from the functional integral standing on the R.H.S. of Eq.
(7), let us pass to the integration over the trajectories
$u_\mu(\lambda)=z_\mu(\lambda)+\frac{\lambda}{s}\left(x^\prime-
x\right)_\mu - x_\mu^\prime$, which yields

$$\Delta S=J^2g^2\int d^2\xi\int
d^2\xi^\prime\varepsilon^{ij}\varepsilon^{kl}\left(\partial_ix_\mu\right)
\left(\partial_j x_\nu\right)(\partial_k
x_\rho^\prime)\left(\partial_lx_\nu^\prime\right)\frac{\partial^2}{\partial 
x_\mu\partial x_\rho^\prime}\int\limits_0^\infty ds e^{-\frac{\left(x-
x^\prime\right)^2}{4s}}\int \left(Du\right)_{00}e^{-\int\limits_0^s\frac
{\dot u^2}{4}d\lambda}\cdot$$

$$\cdot\left<tr{\,}P{\,}\exp\left[ig\int\limits_0^s d\lambda
\left(\frac{x-x^\prime}{s}+\dot u\right)_\alpha B_\alpha\left(u+x^\prime+ 
\frac{\lambda}{s}\left(x-x^\prime\right)\right)\right]\right>_{B_\mu^a}, 
\eqno (9)$$
and from now on we shall absorb the inessential constant factors into the 
measure $DB_\mu^a$.

The Wilson loop standing on the R.H.S. of Eq. (9) may be expanded as follows

$$tr{\,}P{\,}\exp\left[ig\int\limits_0^s d\lambda\left(\frac{x-x^\prime}{s}
+\dot u\right)_\alpha B_\alpha\left(u+x^\prime+\frac{\lambda}{s}\left(x-
x^\prime\right)\right)\right]=tr{\,}P{\,}\exp\left(ig\int\limits_0^s d
\lambda\dot u_\alpha B_\alpha(u)\right)+$$

$$+ig{\,}tr\int\limits_0^s d\lambda\left[P{\,}\exp\left(ig\int\limits_0^
\lambda d\lambda^\prime\dot u_\alpha B_\alpha\left(u\left(\lambda^\prime
\right)\right)\right)\right]\Biggl[\frac{\left(x-x^\prime\right)_\beta}
{s}B_\beta\left(u\left(\lambda\right)\right)+$$

$$+\left(x^\prime+\frac{\lambda}
{s}\left(x-x^\prime\right)\right)_\beta\dot u_\gamma\left(\lambda\right)
\left(\partial_\beta B_\gamma\left(u\left(\lambda\right)\right)\right)\Biggr]
\left[P{\,}\exp\left(ig\int\limits_\lambda^s d\lambda^{\prime\prime}\dot u_
\zeta B_\zeta\left(u\left(\lambda^{\prime\prime}\right)\right)\right)\right]+
O\left(g^2\right),$$
and since we are working in the lowest order of perturbation theory, all the 
terms of this expansion except for the first one will be omitted below, so 
that Eq. (9) yields

$$\Delta S=J^2g^2\int d^2\xi\int d^2\xi^\prime\varepsilon^{ij}\varepsilon^
{kl}(\partial_ix_\mu)(\partial_jx_\nu)(\partial_kx_\rho^\prime)(\partial_l
x_\nu^\prime)\cdot$$

$$\cdot\int\limits_0^\infty\frac{ds}{s}\left(\frac{\left(x-x^\prime
\right)_\mu\left(x-x^\prime\right)_\rho}{2s}-\delta_{\mu\rho}\right)
e^{-\frac{\left(x-x^\prime\right)^2}{4s}}\Phi (s), \eqno (10)$$
where

$$\Phi (s)\equiv\int \left(Du\right)_{00}e^{-\int\limits_0^s\frac{\dot 
u^2}{4}d\lambda}\left<tr{\,}P{\,}\exp\left(ig\int\limits_0^s d\lambda
\dot u_\alpha B_\alpha(u)\right)\right>_{B_\mu^a}. \eqno (11)$$

Finally, in order to get the desirable correction to the action (1),
we shall expand the R.H.S. of Eq. (10) in powers of $\frac{s}{r^2}$
(according to the discussion in the paragraph before Eq. (8)),
keeping in this expansion terms not higher in the derivatives w.r.t.
world sheet coordinates than the rigidity, which corresponds to the
expansion up to the second order in the parameter $\frac{s}{r^2}$.
Omitting the full derivative terms of the form $\int
d^2\xi\sqrt{g}R$, where $R$ is a scalar curvature of the world sheet,
we get analogously to Ref. 1 the following values of the integrals
standing on the R.H.S. of Eq. (10)

$$\int d^2\xi\int d^2\xi^\prime \varepsilon^{ij}\varepsilon^{kl}
\left(\partial_ix_\mu\right)\left(\partial_jx_\nu\right)(\partial_k
x_\rho^\prime)\left(\partial_lx_\nu^\prime\right)
\left(x-x^\prime\right)_\mu
\left(x-x^\prime\right)_\rho\int\limits_0^\infty\frac{ds}{s^2}
e^{-\frac{\left(x-x^\prime\right)^2}{4s}}\Phi
(s)=$$

$$=4\pi\int\limits_0^\infty ds\Phi (s)\left(4\int
d^2\xi\sqrt{g}-3s\int d^2\xi\sqrt{g}g^{ij}\left(\partial_i
t_{\mu\nu}\right)\left(\partial_j t_{\mu\nu}\right)\right) \eqno
(12)$$
and

$$\int d^2\xi\int d^2\xi^\prime\varepsilon^{ij}\varepsilon^{kl}\left(
\partial_ix_\mu\right)\left(\partial_jx_\nu\right)(\partial_k
x_\mu^\prime)\left(\partial_lx_\nu^\prime\right)\int\limits_0^\infty
\frac{ds}{s}e^{-\frac{\left(x-x^\prime\right)^2}{4s}}\Phi
(s)=$$

$$=2\pi\int\limits_0^\infty ds\Phi (s)\left(4\int
d^2\xi\sqrt{g}-s\int d^2\xi\sqrt{g}g^{ij}\left(\partial_i
t_{\mu\nu}\right)\left(\partial_j t_{\mu\nu}\right)\right). \eqno
(13)$$
Combining together Eqs. (12) and (13), we arrive at
the following correction to the effective action (1) due to the
accounting for the perturbative gluons in the lowest order of
perturbation theory

$$\Delta S=\left(\Delta\frac{1}{\alpha_0}\right)\int
d^2\xi\sqrt{g}g^{ij}\left(\partial_i
t_{\mu\nu}\right)\left(\partial_j t_{\mu\nu}\right), \eqno (14)$$
where

$$\left(\Delta\frac{1}{\alpha_0}\right)=J^2g^2\int\limits_0^\infty
dss\Phi (s). \eqno (15)$$
Notice, that as it was already pointed out in the Introduction,
perturbative gluons do not change the value of the string tension (2)
of the Nambu-Goto term and affect only the coupling constant of the
rigidity term. Since this correction (15) to the nonperturbative
rigid string coupling constant (3) is a pure perturbative effect, its
sign is unimportant for the explanation of confinement in terms of
the dual Meissner effect (see discussion in the Introduction), which
allowed us to refer the constant factor in Eq. (15) to the measure
$DB_\mu^a$ in Eq. (11). The nontrivial content of this correction
emerges due to the path integral defined by Eq. (11), which includes
the background Wilson average.

\vspace{6mm}
{\large \bf 3. Conclusion}

\vspace{3mm}
In this letter we have applied perturbation theory in the
nonperturbative QCD vacuum$^{9}$ and the non-Abelian Stokes theorem,
which represents a Wilson loop in the $SU(2)$ gluodynamics as an
integral over all the orientations in colour space$^{16}$ to the
derivation of the correction to string effective action (1), found in
Ref. 1, which emerges due to accounting for the interaction of
perturbative gluons with the string world sheet in the lowest order
of perturbation theory. This correction is given by formulae
(11), (14) and (15) and affects only the rigidity term, while the string
tension of the Nambu-Goto term keeps its pure nonperturbative value
(2). Perturbative correction (15) to the inverse coupling constant of
the rigidity term contains the dependence on the background fields in
the form of the background Wilson average standing under a certain
path integral (11).

We have also demonstrated that perturbative fluctuations, when being
taken into account, lead to the interaction defined by the R.H.S. of
Eq. (7) between elements of the string world sheet by virtue of
nonperturbative gluonic strings, which agrees with the qualitative
scenario of excitation of the gluodynamics string by the perturbative
gluons, suggested in Refs. 1 and 10.

However, it is still remains unclear whether perturbative gluons may
yield cancellation of the conformal anomaly in $D=4$ rather than in
$D=26$, as it takes place for the ordinary bosonic string 
theory$^{18,19}$, and the
solution of the problem of crumpling for the rigidity term$^{19,20}$.
These problems will be treated in the next publications.

\vspace{6mm}
{\large \bf 4. Acknowledgments}

\vspace{3mm}
I am deeply grateful to Professor Yu.A.Simonov for a lot of useful
discussions on the problem studied in this paper. I would also like
to thank the theory group of the Quantum Field Theory Department of
the Institut f\"ur Physik of the Humboldt-Universit\"at of Berlin
for kind hospitality.

\vspace{6mm}
{\large \bf References}

\vspace{3mm}
\noindent
1.~D.V.Antonov, D.Ebert and Yu.A.Simonov, {\it Mod.Phys.Lett.} {\bf
A11}, 1905 (1996) (preprint DESY 96-134).\\
2.~M.B.Halpern, {\it Phys.Rev.} {\bf D19}, 517 (1979); I.Ya.Aref'eva,
{\it Theor.Math.Phys.} {\bf 43}, 111 (1980); N.Brali\'c, {\it
Phys.Rev.} {\bf D22}, 3090 (1980).\\
3.~Yu.A.Simonov, {\it Yad.Fiz.} {\bf 50}, 213 (1989).\\
4.~N.G. Van Kampen, {\it Stochastic Processes in Physics and
Chemistry} (North-Holland Physics Publishing, 1984).\\
5.~M.Campostrini, A. Di Giacomo and G.Mussardo, {\it Z.Phys.} {\bf
C25}, 173 (1984).\\
6.~A. Di Giacomo and H.Panagopoulos, {\it Phys.Lett.} {\bf B285},
133 (1992).\\
7.~I.-J.Ford et al., {\it Phys.Lett.} {\bf B208}, 286 (1988);
E.Laermann et al., {\it Nucl.Phys.} {\bf B26} (Proc. Suppl.), 268
(1992).\\
8.~ H.G.Dosch, {\it Phys.Lett.} {\bf B190}, 177 (1987); Yu.A.Simonov,
{\it Nucl.Phys.} {\bf B307}, 512 (1988); H.G.Dosch and Yu.A.Simonov,
{\it Phys.Lett.} {\bf B205}, 339 (1988), {\it Z.Phys.} {\bf C45}, 147
(1989); Yu.A.Simonov, {\it Nucl.Phys.} {\bf B324}, 67 (1989), {\it
Phys.Lett.} {\bf B226}, 151 (1989), {\it Phys.Lett.} {\bf B228}, 413
(1989), {\it Yad.Fiz.} {\bf 54}, 192 (1991); H.G.Dosch, A. Di Giacomo
and Yu.A.Simonov, in preparation.\\
9.~Yu.A.Simonov, {\it Yad.Fiz.} {\bf 58}, 113, 357 (1995), preprint
ITEP 37-95; E.L.Gubankova and Yu.A.Simonov, {\it Phys.Lett.} {\bf
B360}, 93 (1995); Yu.A.Simonov, {\it Lectures at the 35-th
Internationale Universit\"atswochen f\"ur Kern- und Teilchenphysik,
Schladming, March 2-9, 1996} (Springer-Verlag, 1996); A.M.Badalian
and Yu.A.Simonov, {\it Yad.Fiz.} {\bf 60}, 714 (1997); D.V.Antonov,
{\it Yad.Fiz.} {\bf 60}, 365 (1997) ({\it hep-th}/9605044).\\
10. Yu.A.Simonov, {\it Nuovo Cim.} {\bf A107}, 2629 (1984).\\
11. D.V.Antonov and Yu.A.Simonov, {\it Int.J.Mod.Phys.} {\bf A11},
4401 (1996); D.V.Antonov, {\it JETP Lett.} {\bf 63}, 398 (1996), {\it
Mod.Phys.Lett.} {\bf A11}, 3113 (1996) ({\it hep-th}/9612005), {\it
Int.J.Mod.Phys.} {\bf A12}, 2047 (1997), {\it Yad.Fiz.} {\bf 60}, 553
(1997) ({\it hep-th}/9605045).\\
12. P.Orland, {\it Nucl.Phys.} {\bf B428}, 221 (1994).\\
13. S.Mandelstam, {\it Phys.Lett.} {\bf B53}, 476 (1975); G.'t Hooft,
in {\it High Energy Physics} (Ed. A.Zichichi) (Editrice Compositori,
1976).\\
14. D.V.Antonov, {\it Pis'ma v ZhETF} {\bf 65}, 673 (1997) 
({\it hep-th}/9612109).\\
15. A.Yu.Dubin, A.B.Kaidalov and Yu.A.Simonov, {\it Yad.Fiz.} {\bf
56}, 213 (1993), {\it Phys.Lett.} {\bf B323}, 41 (1994);
E.L.Gubankova and A.Yu.Dubin, {\it Phys.Lett.} {\bf B334}, 180
(1994), preprint ITEP 62-94.\\
16. D.I.Diakonov and V.Yu.Petrov, in {\it Nonperturbative Approaches
to QCD, Proceedings of the International Workshop at ECT*, Trento,
July 10-29, 1995} (Ed. D.I.Diakonov) (PNPI, 1995), {\it
hep-th}/9606104.\\
17. B.S. De Witt, {\it Phys.Rev.}, {\bf 162}, 1195, 1239 (1967);
J.Honerkamp, {\it Nucl.Phys.} {\bf B48}, 269 (1972); G.'t Hooft, {\it
Nucl.Phys.} {\bf 62}, 444 (1973); L.F.Abbot, {\it Nucl.Phys.} {\bf
B185}, 189 (1981).\\
18. A.M.Polyakov, {\it Phys.Lett.} {\bf B103}, 207 (1981).\\
19. A.M.Polyakov, {\it Gauge Fields and Strings} (Harwood Academic
Publishers, 1987).\\
20. A.M.Polyakov, {\it Nucl.Phys.} {\bf B268}, 406 (1986).
\end{document}